\documentclass{PoS}

\title{The MICE PID Instrumentation}

\ShortTitle{The MICE PID Instrumentation}

\author{\speaker{M.BONESINI}%
         \thanks{\bf{on behalf of the MICE Collaboration}}\\
        Sezione INFN Milano Bicocca\\
        E-mail: \email{maurizio.bonesini@mib.infn.it}}

\abstract{The international Muon Ionization Cooling Experiment (MICE) will carry
out a systematic investigation of ionization cooling of a muon beam.
As the emittance measurement will be done on a particle-by-particle
basis, sophisticated beam instrumentation is needed to measure
particle coordinates and timing vs RF. 
A PID system based on 
three time-of-flight stations, two Aerogel Cerenkov detectors and a KLOE-like calorimeter
has been constructed 
in order to keep beam contamination ($e, \pi$) well below $1 \%$.  
The
MICE time-of-flight system will measure timing with a resolution better than 70 ps
per plane, in a harsh environment due to high particle rates, fringe
magnetic fields and electron backgrounds from RF dark current.}

\FullConference{10th International Workshop on Neutrino Factories, Super beams and Beta beams\\
		 June 30 - July 5 2008\\
		 Valencia, Spain}

\begin{document}

\section{Introduction}
\begin{figure}
\vskip -0.9cm
\begin{center}
\includegraphics[width=.60\linewidth]{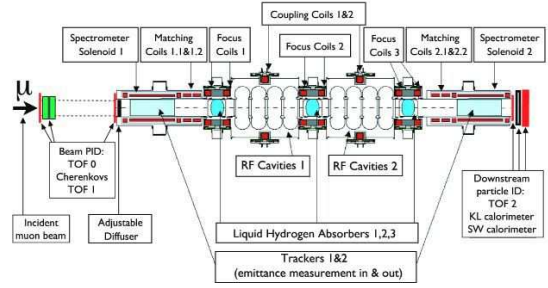}
\end{center}
\caption{2-D layout of the MICE experiment at RAL.
The beam enters from the left. The cooling section
is put between two magnetic spectrometers and two TOF stations
(TOF1 and TOF2)
to measure particle parameters.}
\label{fig:mice}
\end{figure}

The neutrino factory ($\nu F$)~\cite{kosharev} is a muon storage ring with long straight
sections, where decaying muons produce collimated neutrino beams of well defined
composition and high intensity.
Several $\nu$F designs have been proposed, such as the ones of references
\cite{US2,cern}.
A high intensity beam accelerated by a high power Proton Driver 
produces in a Hg jet target, 
after some accumulation and bunch compression, low energy pions.
After a collection system, muons are cooled and phase rotated before acceleration up to 20 GeV/c. Accelerated muons of well defined charge and momentum are then
injected into an accumulator where they circulate until they decay, giving two
neutrino beams along the straight sections.
The physics program at a neutrino factory is very rich and
includes long-baseline $\nu$
oscillations, short-baseline $\nu$ physics and slow muon physics \cite{physrep}.
\noindent
The physics performance of a Neutrino Factory depends not only on its
clean beam composition ($50 \% \nu_{e}, 50 \% \overline{\nu}_{\mu}$ for
the $\mu^{+} \mapsto \overline{\nu}_{\mu} \nu_e e^{+}$ case), but
also on the available beam intensity. The cooling of muons
(accounting for $ \sim 20 \%$ of the final costs of the factory)
is thus compulsory,
increasing the performance up to a factor 10. Due to the short
muon lifetime ($2.2 \ \mu$s), novel methods such as the
ionization cooling, proposed more than 20 years ago by A.N. Skrinsky \cite{skrinsky}, must be used.
Essentially the cooling of the transverse phase-space coordinates of a muon
beam can be accomplished by passing it through an energy-absorbing material
and an accelerating structure, both embedded within a focusing magnetic
lattice. Both longitudinal and transverse momentum are lost in the absorber
while the RF-cavities restore only the longitudinal component.
\noindent
The MICE experiment \cite{mice} at RAL aims at a
systematic study of a section of a cooling channel (see figure \ref{fig:mice} for a
layout).
A secondary muon beam from ISIS (140-240 MeV/c central momentum,
tunable between 1-10 $\pi \cdot $ mm rad input emittance)
enters a cooling section after a diffuser.
The 5.5 m long cooling section consists of three absorbers and eight RF
cavities encircled by lattice solenoids.
The cooling process will be studied by varying the relevant parameters,
to allow the extrapolation to different cooling channel designs.

\section{Emittance measurement}
Different $\nu$F designs need a muon cooling factor from 2 to 16,
over a distance of $\sim 100$ m. The different  designs of the cooling
section  have to be
tested at least with a precision of $\sim 10 \%$. For an affordable
cooling section prototype a cooling $\sim 5-20 \%$ may be exptected.
This implies emittance measurements at a level of $0.1 \%$,
thus excluding conventional emittance measurement methods, with 
precisions around $10 \%$.
To obtain such a pecision a method based on single particle measurements 
has been envisaged.

Particles are measured before and after the cooling section
by two magnetic spectrometers complemented by TOF detectors.
For each particle x,y,t,
x'=dx/dz=$p_x/p_z$,y'=dy/dz=$p_y/p_z$, t'=dt/dz =$E/p_z$ coordinates
are measured.
In this way, for an ensemble of N particles, the
input and output emittances may be  determined  with high precision.
\section{The MICE PID system}
The driving design criteria of the MICE detector system
 are robustness,
in particular of the tracking detectors, to sustain the severe
background conditions in the vicinity of RF cavities and redundancy in PID
in order to keep contaminations ($e, \pi$) well below $1 \%$.
Precision timing measurements are required
to relate the time of a muon to the phase of the RF and simultaneously
for particle identification by time-of-flight (TOF).
A time resolution around 70 ps ($\sigma_t$) provides both effective ($99 \%$)
rejection of beam pions and adequate ($5^{\circ}$) precision of the RF phase.
Particle identification (PID) is obtained upstream the first solenoid by two
TOF stations (TOF0/TOF1) and two threshold Cerenkov counters (CKOVa/CKOVb),
that will provide $\pi/\mu$ separation up to 300 MeV/c.
\noindent
Downstream the PID is obtained via a further TOF station (TOF2)
and a calorimeter (EMCAL), to separate muons from decay electrons.

Due to the large range of beam momenta, it is not possible to
select a single material  
as Cerenkov radiator
that is sensitive to muons and blind to pions over the entire range. 
The chosen solution is  
two different aerogel counters with refractive indices 
1.07 and 1.12, each equipped respectively with four 8" low background EMI9356KA 
PMTs from the earlier Chooz experiment. Thus high purities
are obtained: from $99.7 \%$ with both counters on to $99.98 \%$ with
both counters off in the momentum range between 210 and 365 MeV/c. At
lower momenta, the $\pi/\mu$ separation is obtained by a tof measurement,
both Cerenkov detectors being blind to both particle types.
  
\begin{figure}
\vskip -1cm
\begin{center}
\includegraphics[width=.35\linewidth]{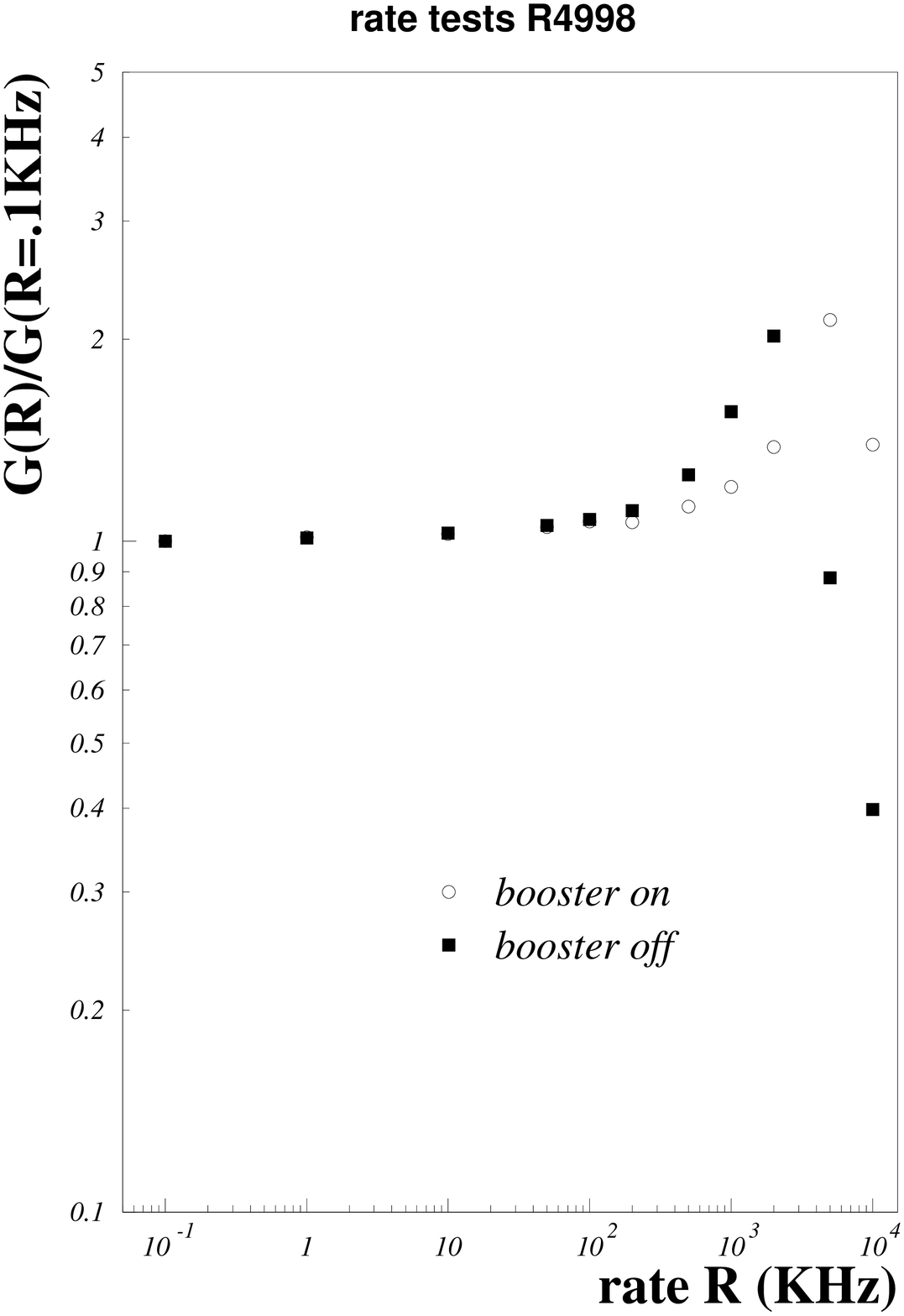}
\includegraphics[width=.60\linewidth]{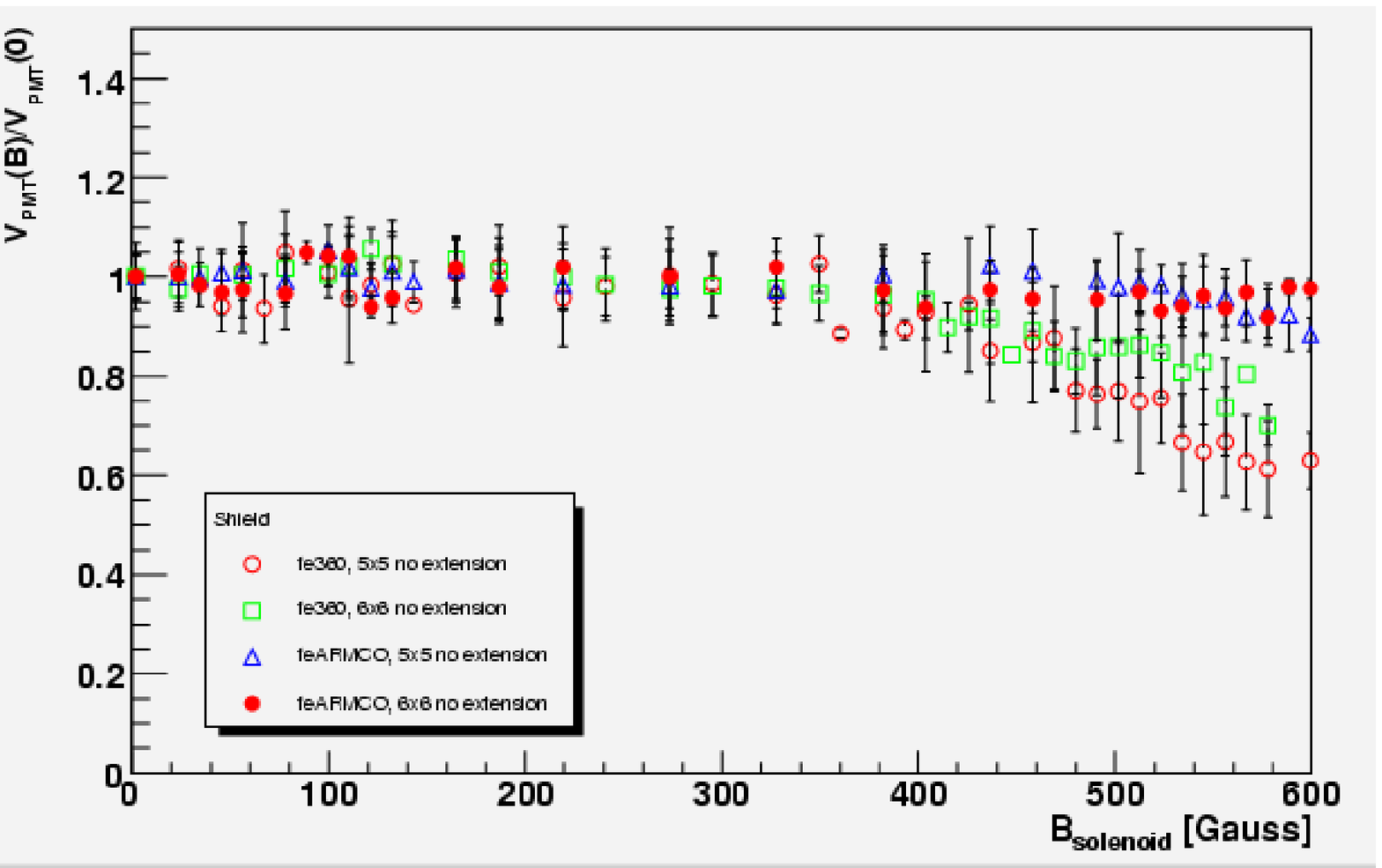}
\end{center}
\caption{Left panel: rate capability of a typical R4998 PMT, as a function of
rate R at field B=0 G (measured P.H. in mV versus rate in KHz).
Right panel: tests of shieldings for conventional R4998 Hamamatsu PMTs.
Results are shown for the average and rms of a sample of 10 PMTs, shielded
with $5 \times 5$ or $6 \times 6 \ cm^2$ low carbon content (Fe360) 
(open circle or open square) or ARMCO (open triangle or black dot) soft iron
box surrounding the mu-metal of the H6533 assemblies. The B field is along the PMT axis.}
\label{fig:pmt}
\end{figure}

All the  TOF stations share a common design 
based on fast 1" scintillator counters
along X/Y directions (to increase measurement redundancy) read at both
edges by R4998 Hamamatsu photomultipliers~\footnote{1" linear focussed PMTs,
typical gain $G \sim 5.7 \times 10^6$ at B=0 Gauss, risetime 0.7 ns, TTS
$\sim 160$ps, equipped with active/passive divider and a mu-metal shield
extending 3 cm in front of the photocathode (H6533 assemblies)}. 
While TOF0 planes cover a $ 40 \times 40$ cm$^2$ active area,
TOF1 and TOF2 cover respectively a $ 42 \times 42$ cm$^2$ and
$60 \times 60$ cm$^2$ active area. The counter width is 4 cm in TOF0 and
6 cm in the following ones.
All downstream
detectors and the TOF1 station must be shielded against stray magnetic
fields (up to 1000-1500 Gauss with a $\leq 400$ Gauss longitudinal component,
depending on the design of the shielding plates after the spectrometer
solenoids). Two options for the local TOF1/TOF2 shielding are foreseen:
in one (for TOF1) a double-sided shielding cage will contain fully the detector,
aside a hole for the beam,   while in the other (for TOF2) individual
massive soft iron box PMTs shielding will be used \cite{ref_D0}.
While the first solution is more elegant and reduces the detector weight,
it gives complications for detector access and maintenance.
Figure \ref{fig:pmt} shows some  results for the shielding of
the most dangerous component of the B field, along the PMT axis, obtained
with mu-metal+a massive iron box shielding. With ARMCO soft iron a good shielding
is obtained up to $\sim 600$ G.

\noindent

The TOF stations must sustain a high instantaneous 
incoming particle rate (up to 1.5 MHz
for TOF0).
PMT rate capabilities were tested in the laboratory with a dedicated
setup \cite{ref_laser} based on a fast laser.
A typical R4998 PMT had a good
rate capability for signals comparable to an incident $\mu$ ($\sim \ 300 \ p.e.$)
up to $\sim 1$ MHz. The rate capability may be increased by the use of active
bases or a booster on the last dynodes for the R4998 PMTs, as shown
in figure \ref{fig:pmt}.

Counter prototypes have been tested at the LNF $DA \Phi NE$ Beam Test Facility
(BTF) with incident electrons of $E_{beam}=200-350$ MeV to study the
intrinsic counter time resolution. The front end readout
used the baseline MICE choice for the TDC: a multihit/multievent CAEN
V1290 TDC,
in addition to a  CAEN V792 QADC (to be replaced in the experiment by
CAEN V1724 FADC) for time-walk corrections. The PMT signal was split
by a passive splitter followed by a leading-edge discriminator before
the TDC line.
An intrinsic single counter resolution $\sim 45-60$ ps was obtained
depending upon beam conditions and the design of lightguides or the
scintillator chosen
(Bicron BC404 or BC420~\footnote{ risetime 0.7 (0.5) ns,
$\lambda^{max}_{emission}=408 (391)$ nm, $\lambda^{bulk}_{att}=160 (110)$ cm
for BC404 (BC420)}, Amcrys-H UPS95F).
With a gaussian fit for the pulse-height distributions,
the number of photoelectrons per single impinging
electron ($N_{pe}$) was estimated. 
From $(<R>/\sigma_R)^2$, where $<R>$ is the peak of the
gaussian and $\sigma_R$ its width, an estimate in the range
200-300 p.e.  for BC420 was obtained, depending on the impact point.
Clearly, this estimation neglects electronic noise and is affected
by the bad (good) scintillator-PMT coupling.

 The downstream calorimeter (EMCAL) is a Pb-scintillating fiber calorimeter (KL), of the KLOE type
 \cite{kloe}, with 1-mm diameter blue scintillating fibers glued between
 0.3 mm thick grooved lead plates
  followed by a muon ranger (SW), made
 of a 1 m$^3$ fully sensitive segmented scintillator block.
 This ``spaghetti'' design for KL offers the possibility of a fine sampling and
 optimal lateral uniformity.
The expected resolution $\sigma_{E} \simeq
5 \%/E$ is fully dominated by sampling fluctuations and is
 linear for electrons or photons in the range 70-300 MeV.
SW will be made with extruded scintillator bars with WLS fibers readout.
A prototype has been recently tested at CERN with good results. In the EMCAL
while KL will measure electrons, SW will measure precisely the muon range.
\section{MICE installation and data taking}
MICE will be accomplished in steps, corresponding
mainly to first characterizing the incoming beam and demonstrating 
the capability
to do a high precision measure of emittance and then
to measure the cooling
for a variety of experimental situations.
\begin{figure}
\begin{center}
\includegraphics[width=.40\linewidth]{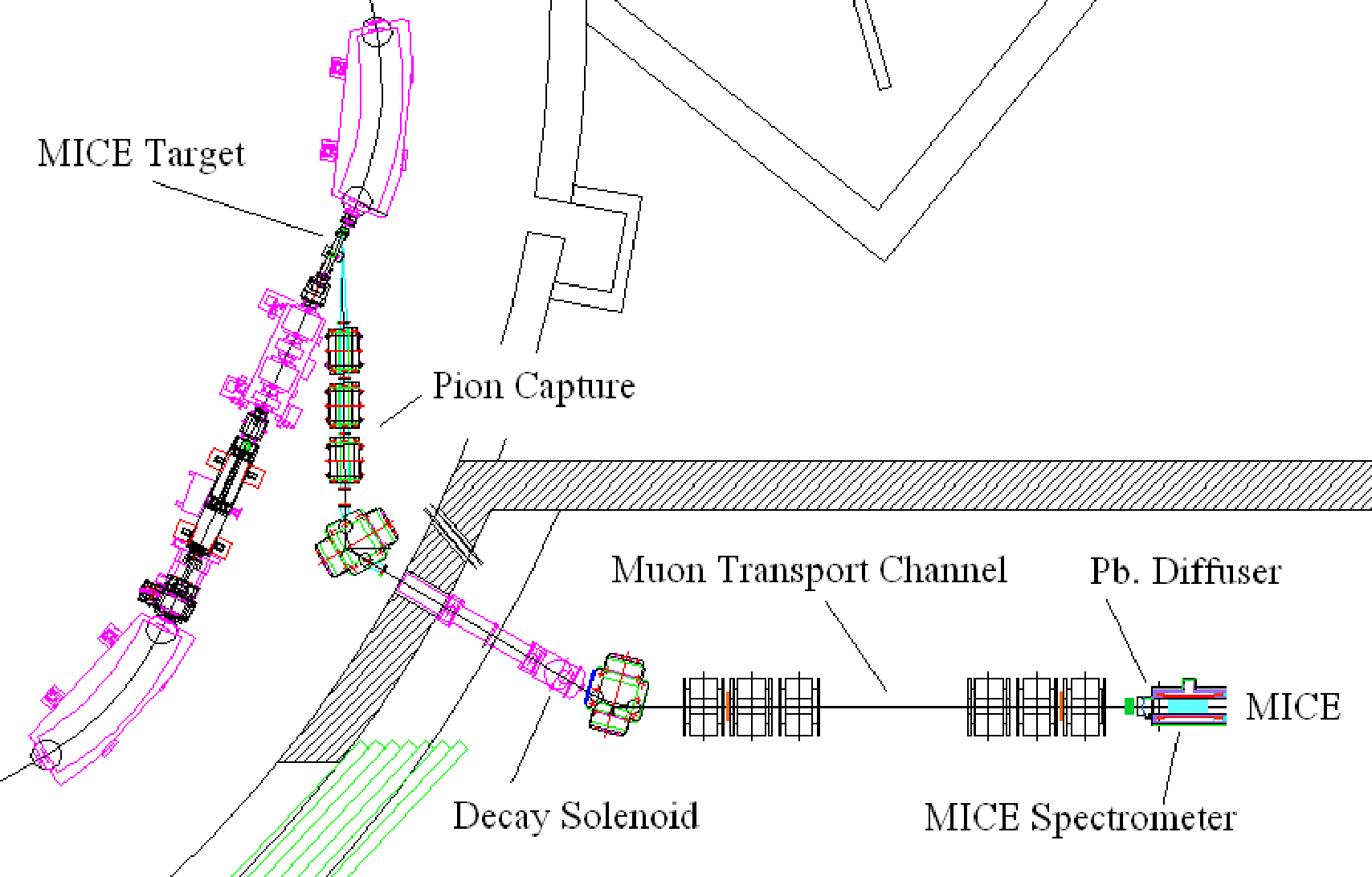}
\includegraphics[width=.40\linewidth]{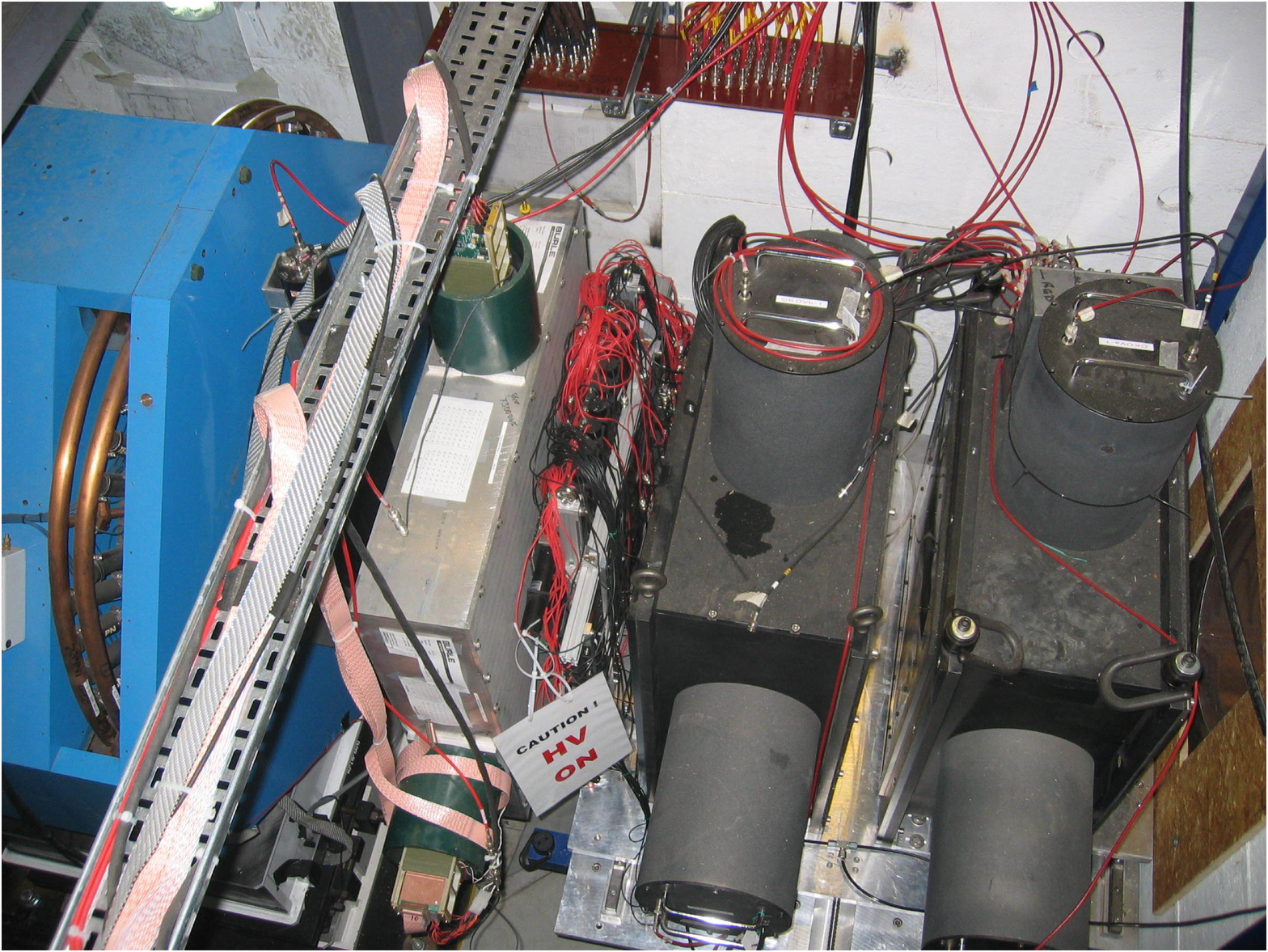}
\end{center}
\caption{Left: layout of the MICE secondary beamline. 800 MeV/c protons 
from ISIS impinge on a Ti target. Produced pions are then guided to a 
decay solenoid, where muons are then steered to the MICE apparatus.The upstream detectors
TOF0 and CKOVa/b are inserted between the two triplets of quads.
Right: picture of the installed TOF0,CKOVa,CKOVb upstream detectors after the
first quads triplet.}
\label{fig:beam}
\end{figure}
Up to now, only the first phase is under way: the beamline has been completed
and the upstream PID detectors and KL installed, see figure \ref{fig:beam}.


\begin{thebibliography}{99}
\bibitem{kosharev} D.G. Kosharev, CERN/ISR-DI/74-62 (174).
\bibitem{US2} M.M. Alsharo'a et al., Phys. ReV. ST. Accel. Beams 6,081001 (2003).
\bibitem{cern} A. Blondel et al., CERN-2004-002.
\bibitem{physrep} M. Bonesini, A.Guglielmi Phys. Rep. 433 (2006) 65.
\bibitem{skrinsky} A.N. Skrinsky, V.V. Parkhomchuk Sov. Jour.  Nucl. Phys.
12 (1981) 3.
\bibitem{mice} A. Blondel et al., MICE proposal, RAL, 2004;
G. Gregoire et al., MICE Technical Report, RAL, 2005.
\bibitem{ref_D0} R. Stephens {\it et al.}, D0 note 2706, 1996.
\bibitem{ref_laser} M. Bonesini {\it et al.}, Nucl. Instr. and Meth. A 572 (2007) 465, \\
M. Bonesini {\it et al.}, Nucl. Instr. and Meth. a 567 (2006) 200.
 \bibitem{kloe} A. Aloisio et al., KLOE coll., Nucl. Inst. and Meth. A 494 (2002), 326.

\end{thebibliography}
\end{document}